# Record Index-Bandgap Trade-off: CdPS$_3$ as a High-Index van der Waals Platform for Ultraviolet-Visible Nanophotonics

Maksim R. Povolotskiy[1†], Aleksandr S. Slavich[2†], Georgy A. Ermolaev[2†], Dmitriy V. Grudinin[2], Nikolay V. Pak[2], Ilya A. Zavidovskiy[1], Konstantin V. Kravtsov[1,2], Anton A. Minnekhanov[2], Mikhail K. Tatmyshevskiy[1], Alexander V. Syuy[2], Dmitry I. Yakubovsky[1], Arslan Mazitov[2], Liudmila A. Klimova[2], Adilet N. Toksumakov[2], Alexander V. Melentev[1], Elena Zhukova[1], Davit A. Ghazaryan[1,3], Gleb Tselikov[2], Ivan Kruglov[2], Sergey M. Novikov[1], Andrey A. Vyshnevyy[2], Aleksey V. Arsenin[1,2], Kostya S. Novoselov[4,5,6], Valentyn S. Volkov[2*]

[1]Moscow Center for Advanced Studies, Kulakova str. 20, Moscow, 123592, Russia

[2]Emerging Technologies Research Center, XPANCEO, Internet City, Emmay Tower, Dubai, United Arab Emirates

[3]Laboratory of Advanced Functional Materials, Yerevan State University, Yerevan 0025, Armenia

[4]National Graphene Institute (NGI), University of Manchester, Manchester, M13 9PL, UK

[5]Department of Materials Science and Engineering, National University of Singapore, Singapore, 03-09 EA, Singapore

[6]Institute for Functional Intelligent Materials, National University of Singapore, 117544, Singapore, Singapore

†These authors contributed equally to this work

*Correspondence should be addressed to e-mail: vsv@xpanceo.com

## Abstract

The development of nanophotonics is hindered by a fundamental trade-off between a material's refractive index ($n$) and its electronic bandgap ($E_g$), which severely restricts the choice of materials for short-wavelength applications. This challenge is particularly acute in the visible and ultraviolet (UV) spectra, where high-performance devices require materials that are simultaneously highly refractive and transparent. Here, we report on the van der Waals (vdW) crystal cadmium phosphorus trisulfide (CdPS$_3$) as a solution to this long-standing problem. Through comprehensive optical and structural characterization, we show that CdPS$_3$ possesses an anomalously high in-plane refractive index across the visible spectrum approaching 3 in the near-UV, combined with a wide indirect bandgap. This combination of properties, which circumvents the empirical Moss's law, is validated by first-principles calculations and direct near-field imaging of highly confined waveguide modes. These findings establish CdPS$_3$ as a leading material for UV-visible photonics, opening a new pathway for the development of high-density integrated circuits and metasurfaces.

**Keywords:** cadmium phosphorus trisulfide, high refractive index, optical constants, waveguides, integrated photonics.

# Introduction

The ongoing paradigm shift in nanophotonics from lossy plasmonics to all-dielectric platforms has intensified the search for materials that combine a high refractive index with low optical loss, particularly in the visible and ultraviolet (UV) spectral ranges[1–3]. Progress toward this goal has been restricted by a materials bottleneck: the lack of materials that offer both a very high refractive index for strong light confinement and a wide bandgap for low-loss operation at short wavelengths. This trade-off, described by Moss's law[4], has forced compromises. Silicon offers a high index but is opaque in the visible[5], while wide-bandgap materials like $TiO_2$ and hBN are transparent but have relatively modest refractive indices[6].

Van der Waals (vdW) materials have been explored as a solution[7–12], with TMDCs showing high indices in the near- and mid-infrared or hBN and $GeS_2$ showing UV transparency[6,13]. However, a critical gap exists for a vdW material combining both properties for visible/UV photonics[14]. The metal phosphorus trichalcogenide ($MPX_3$) family has recently emerged as a promising new platform [15–22], with $CdPS_3$, in particular, having been demonstrated in solar-blind UV photodetectors, hinting at its suitable wide bandgap approaching 3.4 eV[23].

However, to truly assess if $CdPS_3$ can solve the material challenge for short-wavelength nanophotonics, a complete and accurate measurement of its broadband optical constants is essential, but it has not been reported yet. Without knowing the full dielectric tensor, it is impossible to accurately model critical performance metrics like waveguide mode confinement, integration density limited by crosstalk[24], or the efficiency of metasurface elements. This leaves a critical gap between the material's initial promise and its practical implementation in the very technologies it could transform, including advanced integrated systems[6,8].

In this paper, we directly addressed this application-critical need. We report a comprehensive measurement of the complete broadband (250–16000 nm) dielectric function of the vdW crystal $CdPS_3$. We reveal that it possesses an exceptionally high refractive index coexisting with a wide transparency window, addressing the materials challenge for short-wavelength photonics. We validate these optical constants through first-principles calculations and near-field imaging of highly confined waveguide modes. We used these optical constants to model the performance of $CdPS_3$-based waveguides, showing their potential for achieving unprecedented integration densities. Our work provided the missing, foundational data required to unlock $CdPS_3$ as an enabling material for a new generation of high-performance, short-wavelength photonic devices.

# Results

**Crystal Structure and Vibrational Response of van der Waals $CdPS_3$**

Cadmium phosphorus trisulfide ($CdPS_3$) is a layered crystal with a monoclinic lattice belonging to the C2/m space group. Its lattice parameters are a = 6.218 Å, b = 10.763 Å, c = 6.867 Å, and β = 107.58°[25]. The unit cell consists of two layers, held together by weak van der Waals (vdW) forces. As demonstrated in Figure 1a, each cadmium (Cd) atom within a layer is coordinated by six sulfur (S) atoms in an octahedral arrangement, while phosphorus (P) atoms form P-P pairs at the center of $P_2S_6^{4-}$ cluster anions. The resulting pattern represents a honeycomb-like lattice of Cd atoms and $P_2S_6$ groups. The lower picture in Figure 1a represents the layered structure of the crystal.

As illustrated in Figure 1b, a plot of calculated static dielectric constant versus experimental optical bandgap shows that $CdPS_3$ fills a critical gap in the available material landscape. It is positioned between TMDCs, which are primarily functional in the IR, and wide-bandgap materials like hBN and $GeS_2$, which stand out in the UV.

The weak vdW forces facilitate easy mechanical exfoliation into thin flakes, down to a few monolayers (see Figure 1c). The number of layers were estimated through comparison of the measured height (1.3 nm) and theoretical unit cell dimensions (0.63 nm in out-of-plane direction). Nevertheless, in this work we focus on properties of bulk (thickness >100 nm) flakes and their potential applications in nanophotonics.

To verify the crystal quality of the $CdPS_3$ flakes, we exfoliated them onto a porous $SiN_x$ membrane and characterized their structure using scanning transmission electron microscopy (STEM). Figure 1d shows a high-angle annular dark-field (HAADF) image of the transferred flake. The observed lattice spacings would correspond to the known crystallographic planes, confirming that the exfoliated flakes retain the bulk crystal structure. The homogeneous quality of the $CdPS_3$ flakes is evidenced by the sharp, regular pattern that would be seen in a selected area electron diffraction (SAED) image from the same flake. We also validated the chemical composition of the $CdPS_3$ flakes via scanning electron microscopy (SEM) combined with energy-dispersive X-ray spectroscopy (EDX), as shown in Figure 1f. Figure 1g represents SEM image of a bulk $CdPS_3$ flake transferred onto a Si substrate and the corresponding EDX maps of Cd, P, and S elements which confirm the uniform distribution of these atoms throughout the flake. The Cd:P:S (59.6%:20.2%:20.2%) atomic ratio match well with the 1:1:3 stoichiometry of $CdPS_3$.

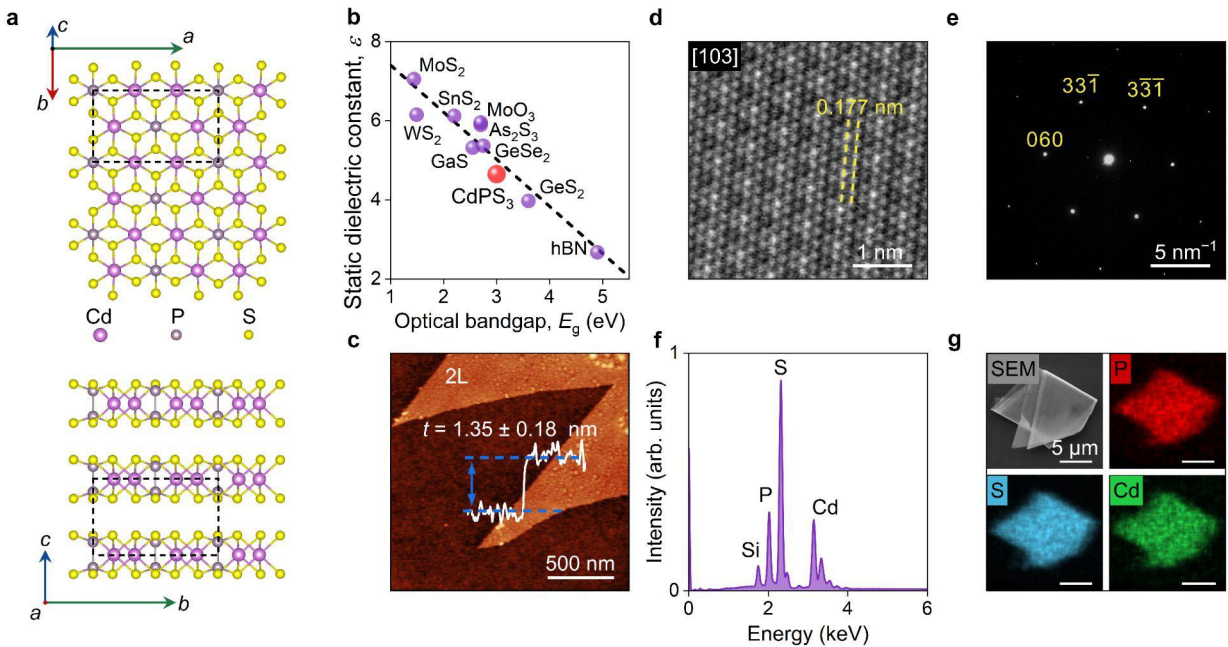

**Figure 1 Crystal structure and characterization of $CdPS_3$. a,** The lattice structure of $CdPS_3$: the top view of an individual monolayer along the [103] direction of the bulk structure and the side view along *a*-axis. **b,** Calculated out-of-plane static dielectric constants of van der Waals materials plotted against the experimental optical bandgaps i.e., the light energy where a material starts absorbing light[13]. For calculation details see Methods. **c,** Atomic force microscopy image of an exfoliated bilayer $CdPS_3$ flake. **d,** A high-resolution HAADF-STEM image of $CdPS_3$ flake. **e,** SAED pattern characteristic of $CdPS_3$ the main reflections associated with reciprocal optical constants. **f,** Energy-dispersive X-ray spectrum (EDX) of the $CdPS_3$ flake transferred onto Si substrate. SEM image of the flake and EDX elemental mapping of corresponding elements: P in red, S in blue and Cd in green.

To investigate the vibrational properties and confirm the crystalline symmetry of the $CdPS_3$ flakes, we performed Raman spectroscopy measurements. The representative Raman spectrum, acquired with a 632.8 nm excitation laser, is presented in the top-left panel of the provided image. The spectrum reveals several distinct phonon modes at Raman shifts of 78, 125, 223, 247, 271, 377 and 564 cm$^{-1}$. The positions and relative intensities of these peaks are in excellent agreement with previously reported spectra for monoclinic $CdPS_3$[26–28], confirming the high quality and purity of the crystal.

Polar plots of the Raman peak intensities for three active modes are shown in Figure 2c–e. The fitting procedure of angular dependences is discussed in Supplementary Note 1. The results reveal both isotropic and anisotropic responses: 247 cm$^{-1}$ and 377 cm$^{-1}$ modes are polarization-independent within experimental uncertainty, whereas 223 cm$^{-1}$ mode exhibits a pronounced anisotropy, consistent with the monoclinic crystal symmetry.

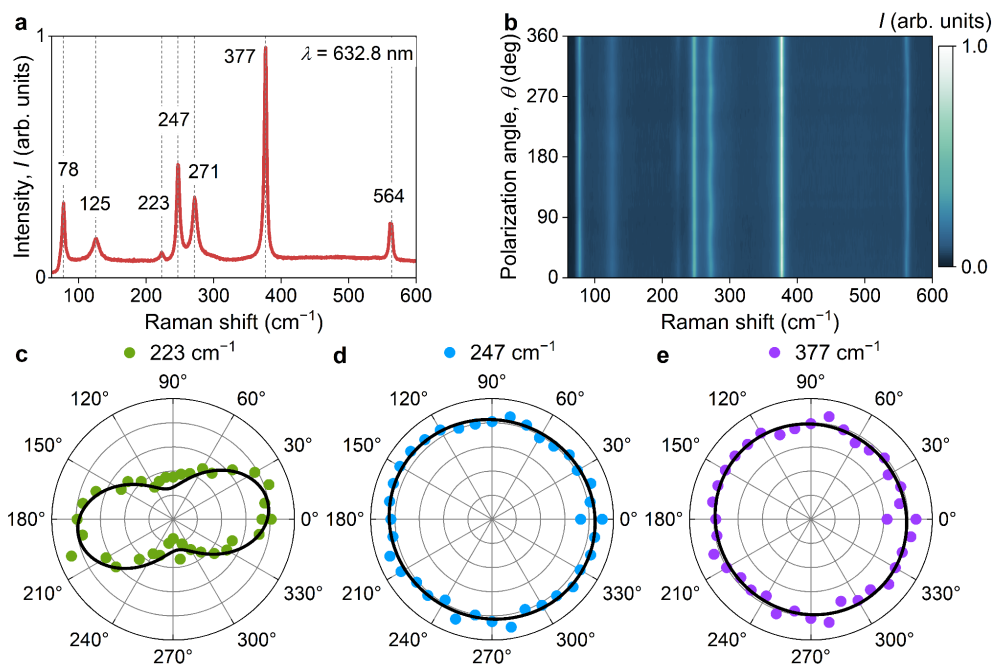

**Figure 2 Polarized Raman spectra of bulk CdPS$_3$. a,** Raman spectrum of bulk CdPS$_3$ at ambient conditions. Dashed lines labelled with numbers denote vibrational mode positions with corresponding wavenumbers in cm$^{-1}$. **b,** Angle-resolved polarized Raman intensity map of bulk CdPS$_3$. **c-e,** Polar plots (color dots) and fittings (dark solid line) of three angle-resolved normalized Raman intensities.

**Broadband Optical Properties of CdPS$_3$**

Despite the structural anisotropy of the monoclinic lattice and pronounced anisotropy of its vibrational modes, cadmium phosphosulfide (CdPS$_3$) unexpectedly exhibits an isotropic in-plane optical response. Figure 3a shows a representative polarization-resolved transmittance map of a CdPS$_3$ flake exfoliated on a Schott glass substrate. Across the visible spectrum (450–850 nm), the transmittance remains virtually independent of the incident light polarization angle.

To retrieve the dielectric constants over an ultrabroad spectral range (250–16000 nm) with high accuracy, we performed a global analysis combining spectroscopic ellipsometry and normal-incidence transmittance and reflectance spectra acquired from multiple flakes of different thicknesses. The full spectroscopic analysis procedure is described in detail in the Supplementary Note 2. As shown in Figure 3b, the representative experimental reflectance $R$ and transmittance $T$ spectra exhibit excellent agreement with calculations based on the extracted optical constants. Figure 3c presents the derived refractive index ($n$) and extinction coefficient ($k$) for both the in-plane (*ab*-plane) and out-of-plane (*c*-axis) directions. The experimental results are strongly supported by *ab initio* calculations, which replicate the dispersion trends.

The in-plane refractive index exceeds $n = 3$ in the visible–near-UV range and decreases gradually towards the IR, while $k_{ab}$ remains near zero. Figure 3d contextualizes these findings by comparing the refractive index of CdPS$_3$ with established high-index materials. CdPS$_3$ surpasses conventional wide-bandgap dielectrics such as TiO$_2$, LiNbO$_3$, and hBN across the visible and near-UV regions. Furthermore, it offers a significantly wider transparency window than GaP or As$_2$S$_3$. This unique combination of a high refractive index ($n$ > 3) and wide bandgap ($E_g$ ≈ 3.4 eV) positions CdPS$_3$ as an exceptional material for short-wavelength nanophotonics, uniquely overcoming the limitations imposed by Moss's rule.

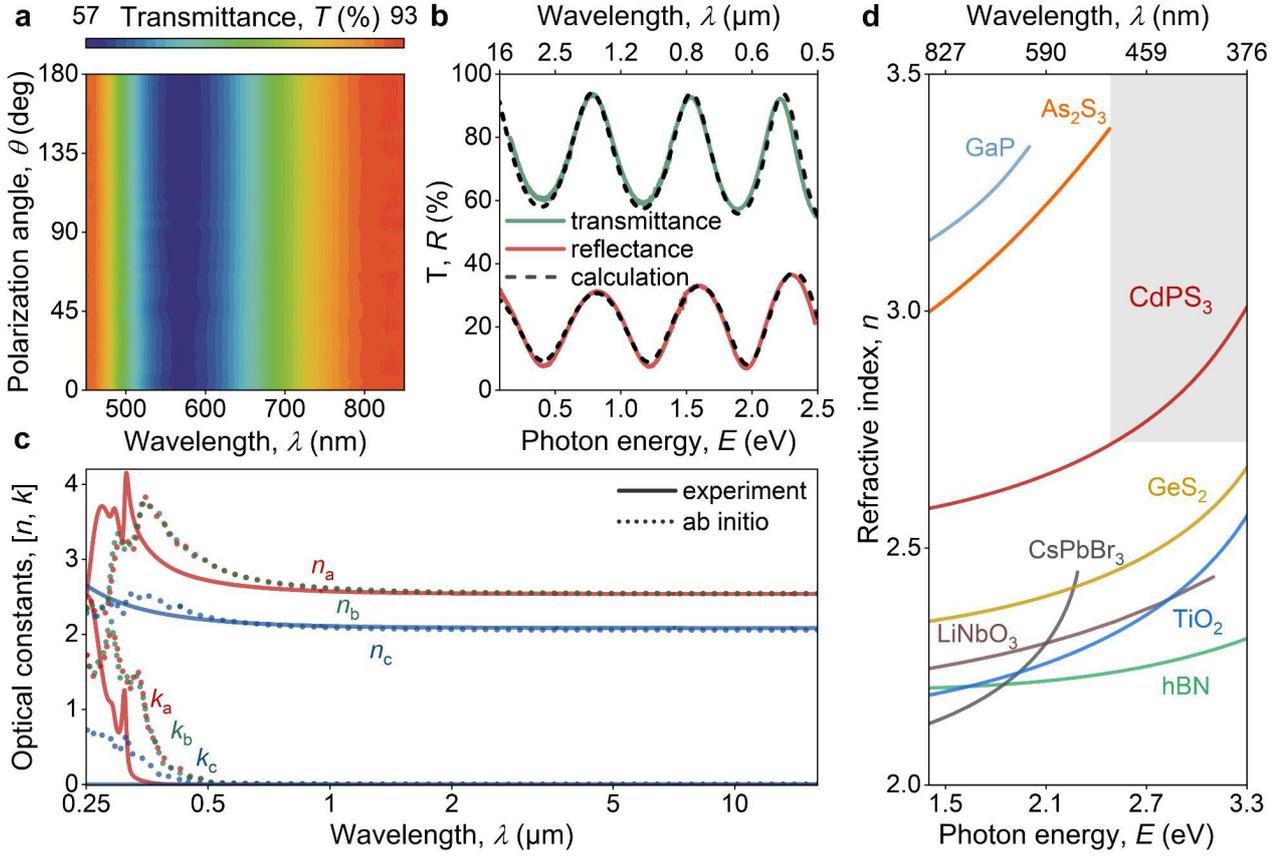

**Figure 3 Optical characterization and comparison of CdPS$_3$ with other high-refractive index materials. a,** Polarization-resolved transmittance map of bulk CdPS$_3$ as a function of wavelength and polarization angle. The color gradient illustrates the transmittance percentage across the spectral range (450–850 nm) **b,** Experimental reflectance $R$ and transmittance $T$ (solid curve) compared to calculated spectra (dashed black curve). **c,** Refractive index $n$ and extinction coefficient $k$ of CdPS$_3$ crystal measured experimentally (solid curves) and calculated using *ab initio* methods (dotted curves). The red (also green for ab *initio*) and blue curves correspond to the in-plane ($n_{ab}$, $k_{ab}$) and out-of-plane ($n_c$, $k_c$) components, respectively. Experimental data were obtained through spectroscopic ellipsometry measurements. **d,** In-plane refractive index of CdPS$_3$ (red curve) compared with high-refractive index materials within transparency window ($k$ < 0.01), including TiO$_2$, GaP[29], hBN[6], LiNbO$_3$[30], CsPbBr$_3$[31], GeS$_2$[13], and As$_2$S$_3$[32]

To validate the experimentally determined optical constants ($n$, $k$) and directly confirm the potential of CdPS$_3$ for strong light confinement, we used scattering-type scanning near-field optical microscopy (s-SNOM). This technique allows for the real-space visualization of guided electromagnetic modes within the CdPS$_3$ flakes.

We investigated flakes of varying thicknesses exfoliated onto CaF$_2$ substrates. When illuminated, light couples into the waveguide modes of the CdPS$_3$ flake. The interference between the background signal and the propagating mode results in distinct fringes (Figure 4a). The periodicity of these fringes corresponds to the waveguide mode momentum. Figure 4b shows line profiles extracted from the near-field maps at $\lambda$ = 700 nm and 1600 nm for a 254-nm-thick-flake. The corresponding Fourier transforms (Figure 4c) reveal sharp peaks corresponding to the effective indices of the transverse electric (TE$_0$) and transverse magnetic (TM$_0$) modes. The effective mode indices $n_{eff}$ are extracted from these momenta after applying a geometric correction[33,34], accounting for the oblique incidence angle of the excitation (see Methods and Supplementary Note 3).

We compared these experimental results with theoretical energy-momentum dispersion relations calculated using the transfer matrix method (TMM)[35], utilizing the optical constants derived from ellipsometry (Figure 3c). The experimental data points for both thin (254 nm, Figure 4d) and thick (1200 nm, Figure 4e) flakes show excellent agreement with the calculated dispersion curves across a broad spectral range (visible to mid-infrared). This agreement rigorously validates the accuracy of the determined dielectric tensor and confirms the strong optical confinement enabled by the high refractive index of $CdPS_3$.

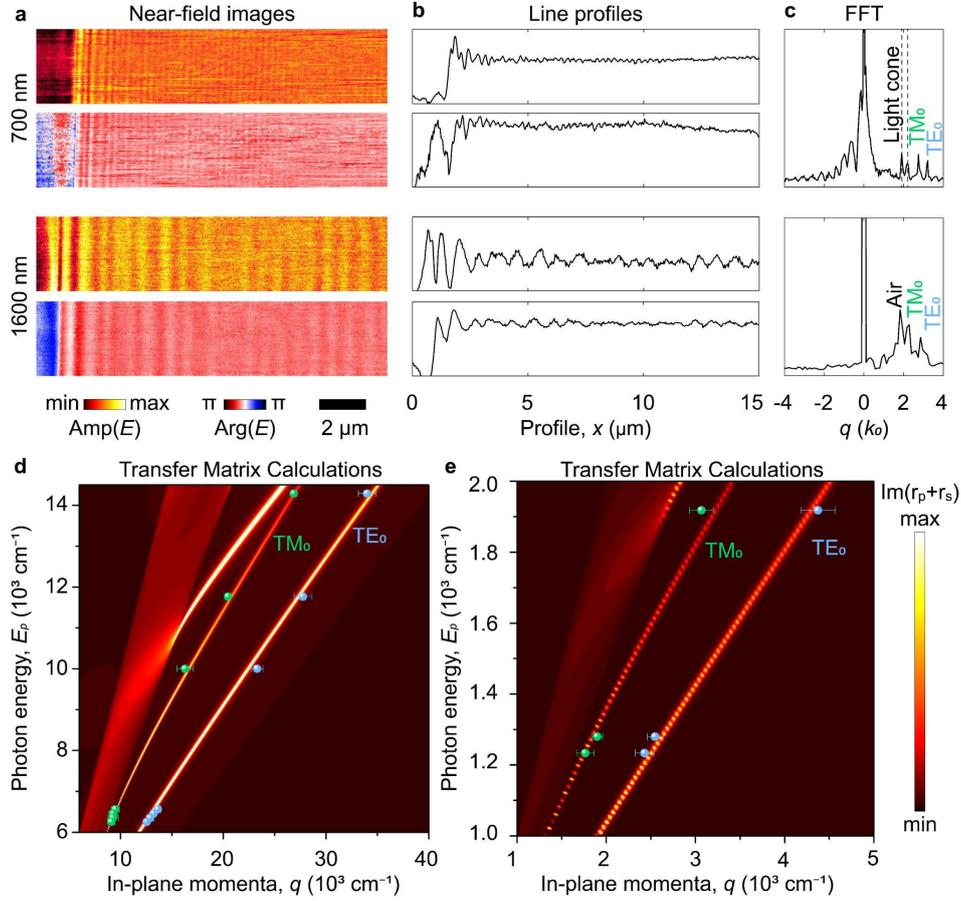

**Figure 4 Near-field analysis of $CdPS_3$. a,** Near-field amplitude Amp(E) and phase Arg(E) maps of a 254 nm flake taken at 700 nm and 1600 nm **b,** x-line scans taken from panel a and averaged over 3 μm along the y-axis. **c,** Complex Fast Fourier Transform (FFT) of the complex near-field signal in (b). **d,** and **e,** Transfer matrix calculations for the Air/$CdPS_3$/$CaF_2$ system for thickness of the $CdPS_3$ 254 nm and 1200 nm respectively. Green and blue dots show data extracted from the near-field experiments for the $TM_0$ and $TE_0$ modes respectively.

**Prospects for High-Density Photonic Integration**

The combination of a high refractive index and low loss at short wavelengths makes $CdPS_3$ a compelling platform for high-density photonic integrated circuits. To quantify this potential, we modeled the performance of $CdPS_3$ waveguides and compared them with conventional photonic materials: $MoS_2$, Si, $TiO_2$, and hBN (see Methods and Supplementary Note 4). We evaluated the performance at the shortest wavelength ($\lambda_{min}$) where each material maintains high transparency, defined here by an extinction coefficient threshold of $k = 0.01$. This corresponds to 375 nm for $CdPS_3$, 370 nm for $TiO_2$, 730 nm for Si, and 865 nm for $MoS_2$. For hBN, which shows no absorption in its reported range, we used the boundary of the available data (250 nm) [6]. We simulated square cross-section waveguides on a $SiO_2$ substrate (Figure 5a).

To establish a fair comparison of confinement, we designed waveguides for each material such that the exponential decay constant ($\chi$) of the evanescent field into the SiO$_2$ cladding was fixed at $\chi$ = 2.49 μm$^{-1}$. The decay constant relates to the effective mode index ($n_{eff}$) and the cladding index ($n_{sub}$) (see Methods). Figure 5b plots the required waveguide width ($w$) versus the operating wavelength ($\lambda_{min}$). Due to its high refractive index, CdPS$_3$ waveguide provides substantially tighter confinement ($w$ = 94 nm at 375 nm) than TiO$_2$ ($w$ = 132 nm at 370 nm). On the other hand, silicon and MoS$_2$, while offering high refractive indices, operate at much longer wavelengths, resulting in larger physical dimensions for the same relative confinement.

The ultimate limit of integration density is determined not only by mode confinement but also by evanescent field coupling (crosstalk) between adjacent waveguides. We calculated the distance required for power transfer between two adjacent waveguides (the crosstalk length, $L$), as a function of the separation distance $d$ (Figure 5c)[24]. The waveguide widths were optimized to minimize crosstalk at each separation distance. The results demonstrate that CdPS$_3$ allows for significantly higher integration densities (shorter $d$ for a given $L$) compared to TiO$_2$ despite CdPS$_3$'s higher operating wavelength. This analysis highlights the capability of CdPS$_3$ to enable sub-100 nm waveguides with high integration density at wavelengths inaccessible to conventional high-index materials.

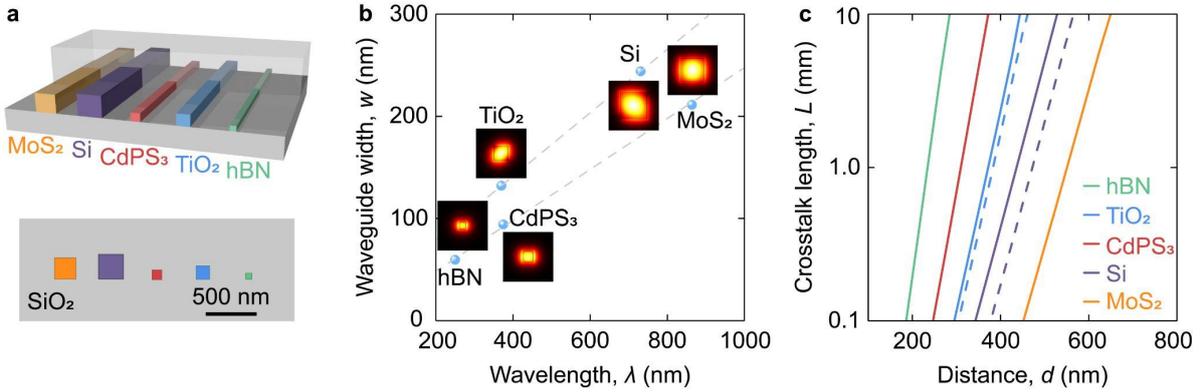

**Figure 5 Prospect applications of CdPS$_3$. a** Schematic comparison of the dimensions of waveguides made from high-refractive index materials: 59×59 nm$^2$ for hBN, 132×132 nm$^2$ for TiO$_2$, 94×94 nm$^2$ for CdPS$_3$, 244×244 nm$^2$ for Si and 211×211 nm$^2$ for MoS$_2$. **b** Waveguide width for different materials with constant $\chi$ = 2.49 mm$^{-1}$. **c** Relation of crosstalk length and distance between cores of the waveguides for high-refractive index materials. The width is optimized for each given distance between waveguides. The dashed lines dashed lines indicate crosstalk for TM modes.

## Conclusion

The search for high-performing photonic materials often feels like a zero-sum game. One could have either a wide transparency window that sacrifices refractive index, or a giant index locked behind the narrow gate of absorption. In this study, we show that this trade-off is not a fundamental roadblock but a material challenge that can be solved. By looking at the MPX$_3$ family we find in CdPS$_3$ a surprisingly elegant solution. Our work provides a comprehensive study on its broadband optical properties, revealing a material that marries a near-ideal high refractive index ($n \approx 3$) with the wide bandgap for lossless operation in the blue part of the visible wavelength range. The immediate consequence is a new roadmap for device miniaturization. The dielectric tensor we report is not merely a catalog of properties; it is the essential blueprint for a new device architecture platform. It allows for a deterministic co-design of material and light, enabling sub-100 nm waveguides and meta-atoms with performance metrics previously thought to be impossible for a transparent material.

## Methods

**Sample preparation.** Bulk $CdPS_3$ single crystals were obtained from 2D Semiconductors Inc. and mechanically exfoliated onto Si, Si/SiO$_2$, and Schott glass substrates at ambient conditions using adhesive scotch tapes (Nitto Denko Corporation). Before exfoliation, the substrates were sequentially treated with acetone, isopropanol, and deionized water, followed by exposure to air plasma to eliminate surface contaminants. For transmission electron microscopy (TEM) characterization, $CdPS_3$ layers were transferred onto porous TEM grids (Norcada) via a dry-transfer process assisted by a polydimethylsiloxane (PDMS) support layer. The thickness and surface morphology of $CdPS_3$ flakes were characterized by an atomic force microscope (NT-MDT Ntegra II) in HybriD mode at ambient conditions.

**Angle-resolved Raman spectroscopy.** Raman spectra were acquired with a Horiba LabRAM HR Evolution (HORIBA Ltd.) confocal microscope with ×100/N.A. = 0.90 objective, parallel analyzer and 1800 lines/mm diffraction grating were used. The studies were carried out at 632.8 nm excitation wavelength. For every polarization angle, Raman mapping was used to obtain an array of 9 spectra measured with 5 μm lateral step. Excitation power density was 125 kW/cm$^2$, integration time was 10 s for each point, spot area was 1.8 μm$^2$. The spectra for each polarization angle were obtained by averaging all spectra of each map. The processing of the spectra was carried out by background subtraction. Afterwards, background-subtracted spectra were fitted by Lorentzian lines. Finally, results were used to fit the data and construct the polar graphs.

**Angle-resolved micro-spectroscopy.** Polarization-resolved transmittance measurements in the visible range were performed using a Zeiss Axio Lab.A1 optical microscope that was illuminated by a halogen light source. The instrument was coupled to an Optosky ATP5020P grating spectrometer using a Thorlabs M92L02 optical fiber with a 200 μm core diameter. Collection of the transmitted light was confined to a spot smaller than 15 μm by an "N-Achroplan" 50× Pol M27 objective lens, itself defined by a numerical aperture of 0.8. For reflection spectra, the same technical solution was employed, but with the use of an RX50M microscope and a SOPTOP MPlanFL, 50×, N.A.=0.8 objective. The complete experimental configuration was detailed previously[36,37]. Spectroscopic measurements were extended into the near- and mid-infrared (NIR/MIR) regions using a Fourier transform infrared spectrometer (Bruker Vertex 80v) that was coupled to an infrared microscope (Hyperion 2000). Both reflectance and transmittance spectra were acquired at a normal incidence geometry using a 15× reflective objective (N.A.=0.4). The NIR band (900–1400 nm) was investigated using a configuration that included a tungsten-halogen lamp source, a calcium fluoride ($CaF_2$) beamsplitter, and a nitrogen-cooled mercury-cadmium-telluride (MCT) detector. To probe the MIR range (1400–17000 nm), a Globar thermal source and a potassium bromide (KBr) beamsplitter were employed.

**Spectroscopic ellipsometry.** Spectroscopic ellipsometry measurements of $CdPS_3$ were performed using spectroscopic imaging ellipsometer Accurion EP4 in the rotating analyzer mode. Ellipsometry spectra were recorded for two flakes of $CdPS_3$ with thicknesses of 50 nm and 165 nm. The spectral range of measurements is 250–1700 nm in 1 nm step and the incident angles of incidence are 45°, 50°, 55°, and 60°.

**Scattering-type scanning near-field optical microscopy.** For near-field measurements we used two different scattering type scanning near-field optical microscopes as well as several different source lasers. Measurements of thick flakes (~1200 nm) were performed in mid-infrared wavelengths using a NTEGRA nanoIR (NT-MDT) scanning near-field optical microscope. To excite the modes in that region we used QD5250C2, QD7500M1 and QD8500CM1 (Thorlabs) quantum cascade lasers working on 5219 nm, 7824 nm and 8912 wavelengths, respectively. We used Pt-coated silicon tip oscillating at the resonance frequency of Ω ≈ 185 kHz with an amplitude of 155 nm (NSG10/Pt). For the measurements of thin flake (~254 nm) in the visible - near-infrared wavelengths region we used a NeaSNOM (Neaspec) s-SNOM. As a sources for that wavelength region we used Ti:Sapphire laser system (TiC, AVESTA) tunable from 700 to 1000 nm, operated at discrete wavelengths of 700 nm, 850 nm, and 1000 nm and a continuous wave Agilent 81600B tunable laser with a tunability range of 1475–1600 nm, operated at six discrete wavelengths with 25 nm step. We also employed various Pt/Ir-coated silicon tips (ARROW-NCPt-50) oscillating at a resonance frequency of approximately $\Omega \approx$ 280 kHz with an amplitude of 70–145 nm. For both wavelength ranges, the microscopes operated in reflection mode, using the same parabolic mirror for both excitation and collection of the near-field signals. To improve the clarity of the near-field images, we reduced the optical background by demodulating the detected signal at a higher-order harmonic frequency $n\Omega$ (with n = 3 in our case), using an interferometric pseudoheterodyne detection scheme with a reference beam modulated by an oscillating mirror.

**Numerical calculations for waveguides.** Numerical calculations for waveguides with fixed decay constant and waveguides crosstalk length were performed using a coupled COMSOL Multiphysics and MATLAB environment via LiveLink. COMSOL was used as a solver, and for every task a simple 2D geometry model was used. MATLAB was used to control simulation parameters and manage the optimization process. To find the exact width of the waveguide for a given $\chi$, we used a binary search algorithm. For optimizing widths of the waveguides for crosstalk calculations we employed a grid-based parameter sweep followed by binary search refinement near the maximum.

**First-principle calculations.** All *ab initio* calculations on optical properties of $CdPS_3$ were done using VASP[38,39]. Lattice parameters and atomic positions were taken from literature[25]. Theoretical results on the frequency-dependent optical constants were obtained within the single-shot GW calculation based on the initial DFT run. Static dielectric constants were calculated within the standard DFT calculation. The kinetic energy with the plane waves basis set (ENCUT) was 400 eV in both steps. The convergence parameter of the self-consistent electronic cycle (EDIFF) was $10^{-8}$ eV. Behavior of the core electrons was described with the projector augmented wave pseudopotentials (POTCAR) of GW type[40]. The exchange-correlation effects were described within the generalized gradient approximation (Perdew–Burke–Ernzerhof functional)[41]. The partial occupancies were set with the Gaussian method (ISMEAR=0) with the smearing (SIGMA) of 0.05 eV. The first Brillouin zone (KPOINTS) was sampled with the 7×7×6 Γ-centered grid. The total number of bands (NBANDS) considered in the both runs was 256. The number of points on the frequency grid in GW calculation (NOMEGA) was 200.

**Transmission electron microscopy.** The samples of $CdPS_3$ were studied on a Titan Themis Z transmission (scanning) electron microscope, which allows studying the fine structure of the samples. The Titan Themis Z microscope is equipped with a sample corrector for correcting spherical aberrations, which significantly improves the resolution of the microscope.

**Energy dispersive X-ray spectroscopy analysis.** The stoichiometry of the $CdPS_3$ flakes crystal flakes was verified by energy dispersive X-ray spectroscopy (EDX), which was performed using a Bruker QUANTAX EDX spectrometer integrated into a scanning electron microscope (SEM, JEOL JSM-7001F) operating in secondary electron imaging mode.

## Acknowledgements

K.S.N. acknowledges support from the Ministry of Education, Singapore, under the Research Centre of Excellence award to the Institute for Functional Intelligent Materials, I-FIM (project No. EDUNC-33-18-279-V12) and under the Tier 3 program (MOE-MOET32024-0001), as well as by the National Research Foundation, Singapore under its AI Singapore Program (AISG Award No: AISG3-RP-2022-028). N.V.P., M.K.T., S.M.N., and A.V.A. acknowledge the financial support from the RSF (Grant No. 25-19-00326).

## Author Contributions

M.R.P., A.S.S., G.A.E. contributed equally to this work. A.V.A., K.S.N. and V.S.V. suggested and directed the project. M.R.P., A.S.S., G.A.E., D.V.G., N.V.P., I.A.Z., A.A.M., M.K.T., A.V.S., D.I.Y., A.V.M., E.Z., G.T. and S.M.N. performed the measurements and analyzed the data. M.R.P., N.V.P., A.N.T., D.A.G. prepared the samples. K.V.K., A.M, L.A.K., I.K., and A.A.V. provided theoretical support. M.R.P., A.S.S., G.A.E. D.V.G. wrote the original manuscript. All authors reviewed and edited the paper.

## Competing Interests Statement

The authors declare no competing interests.